\documentclass[useAMS,usenatbib]{mn2e}

\newcommand{\beq}{\begin{equation}}
\newcommand{\eeq}{\end{equation}}
\newcommand{\beqa}{\begin{eqnarray}}
\newcommand{\eeqa}{\end{eqnarray}}

\newcommand{\ls}{\mathrel{\raise0.27ex\hbox{$<$}\kern-0.70em \lower0.71ex\hbox{{
$\scriptstyle \sim$}}}}

\usepackage{natbib}
\usepackage{enumitem}
\everymath{\displaystyle}
\usepackage{graphicx}
\usepackage{epsfig}
\usepackage{amsmath}
\usepackage{color}
\usepackage{amssymb}
\usepackage[figuresright]{rotating}

\title[Testing Standard Cosmology with Large Scale Structure]{Testing Standard Cosmology with Large Scale Structure}
\author[Stril, Cahn \& Linder]{Arthur Stril$^{1,2}$\thanks{Email: arthur.stril@ens.fr},
Robert N. Cahn$^{2}$
and Eric V. Linder$^{2,3}$  \\
$^{1}$ \'Ecole Normale Sup\'erieure, D\'epartement de Physique, 24, rue Lhomond, 75005 Paris, France\\
$^{2}$Lawrence Berkeley National Laboratory, Berkeley, CA 94720, USA\\
$^{3}$Institute for the Early Universe, Ewha Womans University, Seoul, Korea}

\begin{document}

\date{}

\pagerange{\pageref{firstpage}--\pageref{lastpage}} \pubyear{2009}

\maketitle

\label{firstpage}

\begin{abstract}
The galaxy power spectrum contains information on the growth of structure, 
the growth rate through redshift space distortions, and the cosmic expansion 
through baryon acoustic oscillation features. We study the ability of two 
proposed experiments, BigBOSS and JDEM-PS, to test the cosmological model 
and general relativity. 
We quantify the latter result in terms of the gravitational growth index 
$\gamma$, whose value in general relativity is $\gamma\approx 0.55$. 
Significant deviations from this value could indicate new physics beyond 
the standard model of cosmology.  The results show that BigBOSS (JDEM-PS) 
would be capable of measuring $\gamma$ with an uncertainty 
$\sigma(\gamma) = 0.043$ (0.054), which tightens to 
$\sigma(\gamma) = 0.031$ (0.038) if we include Stage III data priors, 
marginalizing over neutrino mass, time varying dark energy equation of 
state, and other parameters. 
For all dark energy parameters and related figures of merit the two 
experiments give comparable results.  We also carry out some 
studies of the influence of redshift range, resolution, 
treatment of nonlinearities, and bias evolution 
to enable further improvement.
\end{abstract}

\begin{keywords}
Cosmology - Cosmological parameters --- Cosmology - Large-scale structure of the Universe --- Cosmology - Observations
\end{keywords}
  
\section{Introduction}\label{sintro}

Surveys of large-scale structure in the universe provide a rich 
resource for testing our understanding of cosmology. Future 
surveys will cover nearly the full sky to redshifts far deeper 
than are currently studied, mapping out some 10 billion years 
of history.  The great statistical power and 
leverage from 
depth will allow detailed examination of the cosmological 
framework
by carrying 
out a simultaneous fit of a substantial suite of relevant parameters.  
One 
particularly attractive 
prospect is the capability to put to the 
test the predictions of Einstein gravity for the growth of structure 
and its consistency with the cosmic expansion history. 

We consider next-generation surveys mapping the distribution 
of galaxies in three dimensions to redshifts of order 
$z=2$. 
A goal of this study is to determine the 
capabilities of such surveys. 
In particular we aim to 
estimate realistic constraints from a global parameter fit on the 
gravitational growth index $\gamma$, which can characterize 
deviations from general relativity. 
The second goal is to examine how the survey 
characteristics such as redshift range, resolution, and galaxy 
selection affect 
those capabilities.  

In Sec.~\ref{smethod} we review the formalism for extracting 
cosmological information from galaxy correlation measurements in 
terms of the matter power spectrum, and discuss the anisotropic 
distortion due to measuring in redshift space (rather than position 
space).  We discuss the relevant set of cosmological parameters in 
Sec.~\ref{ssolve} and their influence on the matter power spectrum. 
The results are analyzed with emphasis on the role of degeneracies 
between factors that influence growth, including the gravitational 
growth index, the dark energy equation of state, and neutrino mass. 
In Sec.~\ref{ssurvey} we turn to astrophysical and 
survey characteristics and analyze the effect of the bias level of the 
selected galaxy populations, the form of the small-scale velocity 
damping, the spectroscopic survey redshift resolution, and the redshift 
range of the survey.  This allows quantitative comparison of the 
capabilities of next-generation (Stage IV) experiments from both ground 
and space, as well as nearer term (Stage III) experiments.  We conclude 
in Sec.~\ref{sconcl} with a summary of the prospects for testing the 
standard cosmology and revealing clues to dark energy or the breakdown 
of Einstein gravity.

\section{Methodology}\label{smethod}

The future dark energy experiments considered in this paper aim at 
measuring galaxy positions in three dimensions to study 
baryon acoustic oscillations and other aspects of the matter power 
spectrum including its evolution 
through the growth of structure. The matter power spectrum contains 
important cosmological information through its evolving amplitude, its 
shape including the turnover reflecting the transition from radiation 
to matter domination and the suppression due to massive neutrino free 
streaming, and the baryon acoustic oscillation features serving as a 
standard ruler.  

One aspect of particular interest is the 
distorted, anisotropic mapping between the real space density field and 
the measurements in redshift space, caused by peculiar 
velocities \citep{kaiser,redist}. This 
redshift space distortion 
has attracted recent attention as a possible 
technique for detecting deviations from general relativity 
(see \citet{peebgrav,linder-gamma,guzzo} for early work) as it 
depends on the relation between the density and velocity fields, which 
can be altered by 
modifying the gravitational theory. 

Thus the observed galaxy power spectrum contains several types of 
cosmological information.  
The autocorrelation function $\xi(\mathbf{r})$ is defined as the 
excess probability of finding masses at a separation $\mathbf{r}$:

\beqa
  d\mathcal{P} &=& \bar\rho (1+\delta_m) dV\,, \\
  d\mathcal{P}_{12} &=& \bar{\rho}^2 (1 + \xi(\mathbf{r})) dV_1 dV_2\,, 
\eeqa 
where $\bar{\rho}$ is the mean mass density and 
$\delta_m\equiv(\rho-\bar\rho)/\bar\rho$ is the density contrast.  
The mass power spectrum 
is then the Fourier transform of the autocorrelation function:
\beq
P(k) = \int d^3\, \mathbf{r}\, \xi(\mathbf{r})\, e^{i \mathbf{k}\cdot\mathbf{r}}\,, 
\eeq 
with $\mathbf{k}$ the wave vector.  Due to spatial isotropy, only  
the magnitude $k$ will enter.  

We do not observe the power spectrum in real space, however, 
but obtain the radial position through redshift measurements, convolving 
the real distance with additional redshifts due to peculiar velocities.  
This leads 
to the redshift-space power spectrum $\tilde P$ gaining 
an angular dependence through the linear Kaiser factor \citep{kaiser} 
multiplying the isotropic, real space mass power spectrum $P(k)$: 
\beq
\label{eq:full}
\tilde{P}(k,\mu) = (b+f \mu^2)^2 P(k),
\eeq
where $\mu$ is the cosine of the angle that $\mathbf{k}$ makes with 
the line of sight.  For notational simplicity, we suppress the tilde 
from now on.  We work in the linear regime, where the continuity equation 
between the galaxy peculiar velocity field and the galaxy mass overdensity 
is linear (see for example \citet{redist}). 

The dimensionless growth rate $f$ is given by 
\beq
f = \frac{d \ln D}{d \ln a}, \label{eq:fdef} 
\eeq
where $a$ is the scale factor, and $D(a)$ is the growth factor, i.e.\ 
the amplitude $\delta_m(\mathbf{k},a) \propto D(a)$ or $P(k)\propto D^2(a)$. 
We also need to take into account that galaxies, not directly mass density, 
are observed. The bias $b$ relates the galaxy overdensity $\delta_g$ to 
the total mass overdensity through 
$\delta_g = b \delta_m$. 

By looking at the angular dependence of the power spectrum at each $k$,
\beq
P(k,\mu) \propto \sigma_8^2 (b + f \mu^2)^2 = \sigma_8^2 b^2 + 
2 \sigma_8^2 b f \mu^2 + \sigma_8^2 f^2 \mu^4\,, \label{eq:p4} 
\eeq 
where $\sigma_8$ is the normalization of the power spectrum, we can 
in 
principle fit for $b^2 \sigma_8^2$, $b f \sigma_8^2$ and $f^2 \sigma_8^2$, 
hence allowing us to measure $b$ and $f$ provided we have an appropriate 
measurement of $\sigma_8$. This is challenging in practice due to  noise.  
Another possible route to separating out 
the bias involves the 
use of higher order correlation functions \citep{scocc}. 

Although we have three measurable quantities (the three coefficients of 
the fourth order polynomial in Eq.~\ref{eq:p4}) and three unknowns, we cannot 
determine all of them because the second is the geometric mean of the 
other two. This is because we work in the linear regime and general 
relativity, where the galaxy density and peculiar velocity fields are 
perfectly correlated. But should one of these hypothesis be relaxed 
(as in modified gravity models or with non-linearities e.g.\ Finger-of-God 
effects), we need to introduce the correlation coefficient 
between the fields \citep{white,uzan} 
\beq
r(k) = \frac{P_{g v}(k)}{\sqrt{P_{gg}(k) P_{vv} (k)}}\,,
\eeq
where the subscript \textit{g} denotes the galaxy density field, and 
\textit{v} the divergence of the peculiar velocity field.  Ideally this 
correlation would be predicted by the physical theory \citep{dessheth}; 
allowing $r$ 
instead to be completely free significantly degrades the constraints 
on $f$ \citep{white}. We do not consider this situation further in this 
article, instead assuming the standard correlation of unity, since we 
restrict our 
analysis to the linear regime and many classes of gravity theory 
maintain the correlation in this regime. 

To incorporate a measure of the sensitivity to the gravity theory we 
use the gravitational growth index formalism of \citet{growth}, which 
parameterizes 
the growth factor as 
\beq
\label{eq:gro}
D(a) = a\,\exp \left( \int_0^a [\Omega_m(a')^\gamma-1]\,\frac{da'}{a'} \right),
\eeq
so
\beq
f=\Omega_m(a)^\gamma\,,
\eeq
where 
\beq 
\Omega_m(a) = \frac{\Omega_m a^{-3}}{\sum_i \Omega_i \exp \left(3 \int_a^1 \frac{da'}{a'} [1 + w_i(a') ] \right)}
\eeq 
is the ratio of matter density to the total energy density at scale 
size $a=(1+z)^{-1}$. The summation runs over all the different components 
of the universe: matter, dark energy, curvature and radiation. The 
gravitational growth index $\gamma$ will be a parameter of key interest.  
It can distinguish other theories from Einstein gravity -- (see for example 
\citet{growth,cahn-gamma,guzzo}). The merit of a large scale structure 
survey in terms of its gravitational probative power 
may be conveniently 
quantified by the uncertainty $\sigma(\gamma)$. The Figure of Merit 
Science Working Group \citep{StageIII} found that for the suite of 
future Stage III experiments, expected to be completed  
before the proposed Joint Dark Energy Mission (JDEM) program, the  
anticipated uncertainty is $\sigma(\gamma) = 0.21$. 

The standard technique for making such parameter estimation predictions 
is the Fisher matrix \citep{tth}.  For a survey covering a volume $V_0$ 
where the 
mean galaxy number density is $\bar{n}$, the element of the Fisher matrix 
for parameters $p_i$ and $p_j$  is obtained as an integral over the space 
of modes $\mathbf{k}$ \citep{teg}, by: 
\beq
\label{Fish}
F_{ij} = \frac{V_0}{2(2 \pi)^3} \int d^3 k  \left( \frac{\bar{n} P(k,\mu)}{1 + \bar{n} P(k,\mu)} \right)^2 \frac{\partial \ln P}{\partial p_i} \frac{\partial \ln P}{\partial p_j}.
\eeq
The accessible modes are weighted due to shot noise $1/\bar n$ according 
to an effective volume \citep{fkp}: 
\beq
V_{e}(k,\mu) = V_0 \left( \frac{\bar{n} P(k,\mu)}{1 + \bar{n} P(k,\mu)} \right)^2 \,. \label{eq:veff} 
\eeq 
The constraint leverage comes mostly from regions where $\bar{n} P(k,\mu) 
\gtrsim 1$, that is $V_{e} \approx V_0$. 

In order to avoid the uncertainties associated with treatment of 
non-linearities, we truncate the Fisher matrix integral at a 
maximum value $k_{+}$. We take $k_{+} = 0.1\,h$/Mpc, which is the scale 
where departures from linear theory begin to become significant (see, 
e.g., the analysis of \citet{dep}).  See Sec.~\ref{snonlin} for 
a further investigation of non-linear effects. 

The information about $\gamma$ comes from two different parts of the 
power spectrum. The real space, isotropic part, corresponding to no redshift 
space distortions, or $\mu=0$ (observations transverse to the line 
of sight) in the linear regime, is proportional to the growth factor 
squared:
\beq
\label{eq:iso}
P_{\perp}(k) = b^2 P(k) \propto D(a)^2.
\eeq
Note that surveys lacking sufficient redshift resolution are only 
sensitive to the transverse modes due to smearing along the line of 
sight (see, e.g., \citet{nikhil}). 
Using Eq.~(\ref{eq:gro}), the information carried by this part 
involves 
\beq
\frac{\partial \ln P_{\perp}}{\partial \gamma} = 2 \int_0^a \Omega_m(a')^\gamma \ln \Omega_m(a')\, \frac{da'}{a'}.
\eeq
The redshift space distortions in the power spectrum give 
further information through the parameter $f$, which with Eqs.~(\ref{eq:fdef}) 
and (\ref{eq:gro}) reads
\beq
f = \Omega_m(a)^\gamma\,.
\eeq
Therefore, if we define the anisotropic part alone as
\beq
\label{eq:aniso}
P_{\rm aniso}(k,\mu) \equiv 2 b f \mu^2 + f^2 \mu^4  ,
\eeq
it carries information on $\gamma$ 
through 
\beq
\frac{\partial \ln P_{\rm aniso}}{\partial \gamma} = \ln \Omega_m(a) + 
\frac{\Omega_m(a)^\gamma \ln \Omega_m(a) \mu^2}{2b + \Omega_m(a)^\gamma \mu^2}\,.
\eeq 
This factor gives a sense of the information from the redshift distortions. 

Because the measurements become noisier when subdivided into 
angular bins, and because a substantial majority of the information 
resides in the spherically averaged power spectrum \citep{martincom}, 
analyses frequently use the one dimensional, spherically averaged power 
spectrum 
\beq 
P_{\rm sph}(k)=P(k)\left(b^2+\frac{2}{3} b f + \frac{1}{5} f^2 \right)\,. \label{eq:spher} 
\eeq 
This incorporates information from both the original isotropic 
power spectrum and the redshift distortion anisotropies, and may be 
most familiar in terms of the $D_V\propto [D_A^2/H(z)]^{1/3}\propto 
(k_\perp^2 k_\parallel)^{-1/3}$ factor of \citet{eis05}.  In particular, 
the sensitivity to $\gamma$ arises from 
\beq 
\frac{\partial \ln P_{\rm sph}}{\partial \gamma} = \frac{\partial \ln D^2}{\partial \gamma} + 
\frac{[10 b \Omega_m(a)^\gamma + 6 \Omega_m(a)^{2 \gamma}] \ln \Omega_m(a)}{15 b^2 + 10 b\, \Omega_m(a)^\gamma + 3 \Omega_m(a)^{2 \gamma}}\,. 
\eeq 
In Sec.~\ref{ssolve}, we will investigate the relative importance of 
the transverse, anisotropic,  spherically averaged, as well as full 
versions of the power spectrum for constraints on the gravitational 
growth index and other parameters.

\section{Parameter Constraints}\label{ssolve} 

The constraints 
on $\gamma$ 
expected from nearer term (Stage III) surveys are not that 
informative, as mentioned, with $\sigma(\gamma)=0.21$ compared to a 
difference $\Delta\gamma=0.13$ \citep{luess,growth,cahn-gamma} between 
general relativity and DGP gravity \citep{dgp,ddg} for example.  We 
therefore turn to Stage IV experiments and assess their potential for a 
more accurate test of the standard cosmological model.  

We consider two versions of Stage IV power spectrum experiments: 
BigBOSS \citep{bigboss} is a proposed 
ground-based wide field spectroscopic survey and JDEM-PS 
\citep{jdem} is a proposed space-based wide field grism 
survey. 
Both aim at measuring the three 
dimensional spatial distribution of galaxies to study baryon acoustic 
oscillations and the growth of structure. Both experiments 
would use 
the Stage III experiment BOSS \citep{boss}, detecting luminous red galaxies 
(LRG) out to $z=0.7$, as a springboard to higher redshifts. BigBOSS 
would
extend mapping of LRG out to $z=1$ and to the southern sky and 
both experiments 
would
supplement LRG with different classes of emission 
line galaxies (EL) out to $z\approx2$. 

Following \citet{bigboss,jdem,anze}, we give in Table~\ref{spec} the 
redshift range, survey solid angle $\Omega_{sky}$, expected target galaxy 
bias factors $b_{LRG}$ and $b_{EL}$, mean galaxy number density $\bar{n}$, 
and wavelength resolution $R=\lambda/\Delta\lambda$ of the spectrographs 
to be 
used (so the redshift resolution $\sigma_z=\delta z/(1+z)=R^{-1}$). 
We consider variations in redshift, number density, and bias in 
Sec.~\ref{ssurvey}.

\begin{table}
\centering
\begin{tabular}{ccc}
\textbf{BigBOSS} & LRG$^a$ & EL\\
\hline
$z$ range & $0-1$ & $1-2$\\
$\Omega_{\rm sky}$ (deg$^2$) & $24000$ & $24000$\\
$\bar{n}$ ($h$/Mpc)$^3$ & $3.4 \times 10^{-4}$ & $3.4 \times 10^{-4}$\\
$b$ & $1.7$ & $0.8 - 1.2$\\
$R$ & $\geq 2300$ & $\geq 2300$\\
\hline
\hline
\textbf{JDEM-PS} & LRG$^a$ & EL\\
\hline
$z$ range & $0 - 0.7$ & $0.7 - 2$\\
$\Omega_{\rm sky}$ (deg$^2$) & $10000$ & $20000$\\
$\bar{n}$ ($h$/Mpc)$^3$ & $3.4 \times 10^{-4}$ & $19.5 \times 10^{-4}$\\
$b$ & $1.7$ & $0.8 - 1.2$\\
$R$ & $\approx 2000$ & $\geq 200$\\
\end{tabular}
\caption{\label{spec}Survey specifications for the Stage IV experiments 
BigBOSS and JDEM-PS. $^a$Uses northern hemisphere (10000 deg$^2$) LRG 
$z=0-0.7$ from BOSS \citep{boss}.} 
\end{table}

To calculate the power spectrum as a function of redshift and 
cosmological parameters we used the Boltzmann equation code 
CMBeasy \citep{cmbeasy}. 
Using two sided derivatives together with 
convergence tests we can accurately calculate the sensitivity derivatives 
with respect to each parameter.  
These then enter into the Fisher matrix 
calculations of the parameter estimation, taking into account the 
correlations between parameters.  The data points are taken to be the 
power spectrum evaluated at the centers of 10 (or 11) redshift bins 
from $z=0-2$, i.e.\ at $z_i=0.2i+0.1$.  For JDEM-PS, we divide the 
bin containing $z = 0.7$ into two pieces: $z=[0.6,0.7]$ using LRG and 
$z=[0.7,0.8]$ using EL. 

The parameter set involves 9 parameters. Note that when testing the 
gravitational framework, i.e.\ exploring 
beyond-Einstein gravity through quantitative estimation of $\gamma$, it 
is crucial to include all parameters that could act in a similar manner 
on the growth and growth rate.  Therefore we include a time varying 
dark energy equation of state $w(a)=w_0+w_a(1-a)$ and massive neutrinos. 
The parameter list, and the fiducial value around which the Fisher matrix 
expands, is 
\begin{enumerate}[start=1]
  \setlength{\itemsep}{1pt}
	\item $\gamma = 0.55$, gravitational growth index
	\item $b_{LRG}$, the bias for LRG (see Table \ref{spec})
	\item $b_{EL}$, the bias for EL (see Table \ref{spec})
	\item $\Omega_{DE} = 0.744$, dark energy density today
	\item $\Omega_{\nu} = 0.002$, massive neutrino energy density today
  \item $\omega_b = \Omega_b h^2 =0.0227$, reduced baryon energy density today
  \item $h = H_0 / \textrm{(100 km/s/Mpc)} = 0.719$, reduced Hubble constant
	\item $w_0 = -0.99$, dark energy equation of state today 
	\item $w_a = 0$, dark energy equation of state time variation 
\end{enumerate}

The values for $\Omega_{DE}$, $\omega_b$ and $h$ are WMAP-5 best fit 
parameters \citep{wmap}. 
Note that the fiducial $\gamma=0.55$ is the value predicted by General 
Relativity for $\Lambda$CDM (and is quite insensitive to the dark energy 
equation of state); the fiducial $w_0=-0.99$ is taken to avoid issues of 
stepping over $w=-1$.  Dark energy perturbations are included in CMBeasy. 
We assume there is no spatial curvature.  
In the remainder of this section we take the 
fiducial $b_{EL} = 0.8$, and we will investigate the effect of a 
different fiducial in the next section. Note that the neutrino energy density fraction is related to the sum 
of the neutrino masses by $\Omega_{\nu} h^2 =\sum m_{\nu}/\textrm{94 eV}$. 
For a reasonable current upper bound $\sum m_{\nu} \leq 0.3$ eV 
\citep{nubound}, this implies $\Omega_{\nu} \leq$ 0.006.  We take 
$\Omega_{\nu} = 0.002$, or $\sum m_\nu=0.1$ eV as the fiducial. 

Adding together the information from the redshift slices independently 
(note this is not generally a good approximation for slices thinner than 
our $\Delta z=0.2$), we obtain the full Fisher matrix.  We 
do not explicitly add any CMB information (except later when adding 
Stage III Fisher matrices, which assume Planck data). 

Concentrating on testing the gravitational growth index, we now explore 
in more detail what affects the constraints on $\gamma$
using information only from the galaxy power spectrum.  The constraints are 
computed to be 
\beqa 
\sigma(\gamma)_{\rm BigBOSS}&=&0.043 \\ 
\sigma(\gamma)_{\rm JDEM-PS}&=&0.054\,. 
\eeqa 
The importance of including dark energy properties, neutrino masses, 
and other cosmological parameters in the parameter estimation is 
highlighted by the much tighter constraints obtained if we neglect 
their influence, including only $\gamma$ itself and the galaxy biases. 
In this case we obtain 
overly optimistic 
estimates: 
$\sigma(\gamma)_{\rm BigBOSS} = 0.0096$ 
and $\sigma(\gamma)_{\rm JDEM-PS} = 0.0078$.  
Thus, 
taking into account 
the 
correlations with other cosmological parameters is essential.  
The correlation matrices for the 
two experiments are shown in Tables~\ref{cov1} and \ref{cov2}; 
we have replaced the unit diagonal with the uncertainties $\sigma_i$ 
on each parameter. 

\begin{table*}
\centering
$\begin{pmatrix}
\textbf{0.043}  &  -0.50  &  -0.40  &  0.45  &  -0.30  &  0.05  &  0.31  &  -0.93  &  0.88\\
-0.50  &  \textbf{0.021}  &  0.96  &  -0.70  &  0.72  &  -0.09  &  -0.34  &  0.55  &  -0.71\\
-0.40  &  0.96  &  \textbf{0.0099}  &  -0.70  &  0.69  &  -0.08  &  -0.35  &  0.41  &  -0.58\\
0.45  &  -0.70  &  -0.70  &  \textbf{0.0039}  &  -0.12  &  -0.10  &  0.84  &  -0.61  &  0.71\\
-0.30  &  0.72  &  0.69  &  -0.12  &  \textbf{0.0021}  &  -0.29  &  0.33  &  0.18  &  -0.33\\
0.05  &  -0.09  &  -0.08  &  -0.10  &  -0.29  &  \textbf{0.00049}  &  0.01  &  0.00  &  0.01\\
0.31  &  -0.34  &  -0.35  &  0.84  &  0.33  &  0.01  &  \textbf{0.0092}  &  -0.51  &  0.53\\
-0.93  &  0.55  &  0.41  &  -0.61  &  0.18  &  0.00  &  -0.51  &  \textbf{0.16}  &  -0.97\\
0.88  &  -0.71  &  -0.58  &  0.71  &  -0.33  &  0.01  &  0.53  &  -0.97  &  \textbf{0.47}
\end{pmatrix}$
\caption{\label{cov1}BigBOSS correlation matrix for the parameters 
$(\gamma,b_{LRG},b_{EL},\Omega_{DE},\Omega_{\nu},\omega_b,h,w_0,w_a)$. 
The off-diagonal elements are $r_{ij}=C_{ij}/\sqrt{C_{ii} C_{jj}}$ while 
the diagonal elements have been replaced with $\sigma_i=\sqrt{C_{ii}}$ 
in bold.} 
\end{table*}

\begin{table*}
\centering
$\begin{pmatrix}
\textbf{0.054}  &  -0.11  &  -0.33  &  0.33  &  -0.22  &  0.01  &  0.19  &  -0.93  &  0.82\\
-0.11  &  \textbf{0.018}  &  0.91  &  -0.57  &  0.75  &  -0.05  &  -0.10  &  0.26  &  -0.52\\
-0.33  &  0.91  &  \textbf{0.0080}  &  -0.62  &  0.80  &  -0.05  &  -0.11  &  0.41  &  -0.63\\
0.33  &  -0.57  &  -0.62  &  \textbf{0.0028}  &  -0.14  &  -0.14  &  0.76  &  -0.57  &  0.70\\
-0.22  &  0.75  &  0.80  &  -0.14  &  \textbf{0.0019}  &  -0.23  &  0.43  &  0.17  &  -0.37\\
0.01  &  -0.05  &  -0.05  &  -0.14  &  -0.23  &  \textbf{0.00039}  &  0.02  &  0.03  &  -0.02\\
0.19  &  -0.10  &  -0.11  &  0.76  &  0.43  &  0.02  &  \textbf{0.0066}  &  -0.44  &  0.44\\
-0.93  &  0.26  &  0.41  &  -0.57  &  0.17  &  0.03  &  -0.44  &  \textbf{0.14}  &  -0.95\\
0.82  &  -0.52  &  -0.63  &  0.70  &  -0.37  &  -0.02  &  0.44  &  -0.95  &  \textbf{0.37}
\end{pmatrix}$
\caption{\label{cov2}JDEM-PS correlation matrix for the parameters 
$(\gamma,b_{LRG},b_{EL},\Omega_{DE},\Omega_{\nu},\omega_b,h,w_0,w_a)$. 
The off-diagonal elements are $r_{ij}=C_{ij}/\sqrt{C_{ii} C_{jj}}$ while 
the diagonal elements have been replaced with $\sigma_i=\sqrt{C_{ii}}$ 
in bold.} 
\end{table*}

To obtain an overall view of how tightly correlated a parameter is with 
the other variables, we employ the global correlation coefficient -- the 
largest correlation of that parameter with any linear combination of all 
other parameters.
This is given by 
\beq
r_i = \sqrt{1 - \frac{1}{F_{ii}\, (F^{-1})_{ii}}} \,\,. 
\eeq 
We show those vectors in Table~\ref{global}.  Note the high degree of 
correlation, indicating the importance of crosschecks by other data 
and techniques. 

\begin{table}
\centering
$r_{\textrm{BigBOSS}} = \begin{pmatrix}
0.9954\\
0.9943\\
0.9911\\
0.9933\\
0.9993\\
0.9893\\
0.9990\\
0.9997\\
0.9996
\end{pmatrix}$\,\,\, ; \,\,\, $r_{\textrm{JDEM-PS}} =  \begin{pmatrix}
0.9970\\
0.9608\\
0.9960\\
0.9908\\
0.9994\\
0.9895\\
0.9988\\
0.9997\\
0.9996
\end{pmatrix}$
\caption{\label{global}Vectors of the global correlation coefficients 
for the parameters 
$(\gamma,b_{LRG},b_{EL},\Omega_{DE},\Omega_{\nu},\omega_b,h,w_0,w_a)$ 
for BigBOSS and JDEM-PS.}
\end{table}

Examining the marginalized parameter estimations along the diagonals 
of Tables~\ref{cov1} and \ref{cov2}, we see that as expected 
the power spectrum information is especially strong in 
constraining $\Omega_{DE}$ and $h$.  One can determine at the $\sim10\%$ 
level the growth index $\gamma$ and present equation of state $w_0$, 
while $w_a$ and $\Omega_\nu$ have uncertainties of order unity.  
The growth index and equation of state parameters estimation 
is similar for the two experiments: $\gamma=0.55\pm0.043$, $w_0=-0.99\pm0.16$ 
and $w_a=0\pm0.47$ for the ground-based BigBOSS and 
$\gamma=0.55\pm0.054$, $w_0=-0.99\pm0.14$ 
and $w_a=0\pm0.37$ for the space-based JDEM. 
We find the usual high anti-correlation between $w_0$ and $w_a$, and 
a strong correlation between $\gamma$ and $(w_0,w_a)$. 

Regarding the neutrino mass parameter, 
neutrino oscillation experiments indicate 
that neutrinos do have mass \citep{numass,numass2}, but this is not always 
included in parameter estimation despite its correlations.  We 
demonstrate the effect of neglecting this ingredient, finding that it gives 
overly optimistic constraints on $\gamma$ by a factor of three to four.  
The results in Table~\ref{neutr} illustrate the influence of neutrinos 
in three ways, 
including their mass as a free parameter, including their mass but fixing 
its value, and neglecting their mass so they act as a relativistic species. 
At the level of neutrino energy density used as fiducial, $\Omega_\nu=0.002$, 
and over the range $k<0.1\,h$/Mpc used for the power spectrum, the parameter 
value does not strongly affect determination of $\gamma$ and is mainly 
degenerate with the bias parameters. 
However it is crucial to include neutrino mass because the difference 
between treating them as relativistic vs.\ nonrelativistic energy 
density is still important.

\begin{table}
\centering
\begin{tabular}{ccc}
Case & BigBOSS & JDEM-PS\\
\hline
Massive neutrinos, $\Omega_{\nu}$ free & 0.043 & 0.054\\
Massive neutrinos, $\Omega_{\nu}$ fixed & 0.042 & 0.053\\
Relativistic neutrinos & 0.014 & 0.013\\
\end{tabular}
\caption{\label{neutr} Gravitational growth index uncertainty 
$\sigma(\gamma)$ under different treatments of neutrino mass.} 
\end{table}

It is interesting to explore where the main information on the 
gravitational growth index comes from between the transverse and 
anisotropic parts (Eqs.~\ref{eq:iso} and \ref{eq:aniso}), and to compare 
with the spherically averaged case (Eq.~\ref{eq:spher}).  Note that we have 
defined the anisotropic part to isolate the redshift distortion, 
imagining one could remove all shape ($k$) dependence and only focus 
on the angular dependence.  This seems unrealistic and is only 
included as a toy model to highlight 
the $\gamma$ influence on the growth rate $f$; 
the constraints on $\gamma$ become 0.023 for BigBOSS 
and 0.021 for JDEM-PS (note that the parameter space is much reduced, 
with the baryon density, neutrino density, and $h$ not entering). 
Table~\ref{tab:psparts} shows the more realistic parts.

\begin{table}
\centering
\begin{tabular}{ccc}
Case & BigBOSS & JDEM-PS\\
\hline
Transverse (Eq.~\ref{eq:iso}) & 0.126 & 0.128\\
Spherically averaged (Eq.~\ref{eq:spher}) & 0.081 & 0.065\\
Full (Eq.~\ref{eq:full}) & 0.043 & 0.054
\end{tabular}
\caption{Gravitational growth index uncertainty $\sigma(\gamma)$ 
using different parts of the power spectrum.} 
\label{tab:psparts} 
\end{table}

Note the full power spectrum with redshift space distortions has 
the greatest information on the growth index, with a factor 2 better 
constraints than the spherically averaged power spectrum and a factor 
3 better than the transverse (zero redshift distortion or 2D) modes, 
for the BigBOSS case.  BigBOSS achieves these improvements due in large 
part to its high resolution that lets it probe the redshift distortions 
more successfully.  

Finally, a Stage IV power spectrum experiment will not exist in isolation.  
Previous experiments, using several methods, will be carried out 
and the complementarity between methods offers leverage to tighten the 
cosmology constraints.  To study the impact of Stage III priors 
on the parameters we use the Stage III matrix given by the FoMSWG 
website \citep{StageIII} (without double counting the BOSS information), 
rotated into the $(w_0,w_a)$ basis.  
Summing the Fisher matrices of our analysis and of Stage III, we 
extract the constraints on cosmology shown in Table~\ref{pluspre}. 

\begin{table}
\centering
\begin{tabular}{ccc}
 & BigBOSS$_{\rm +III}$ & JDEM-PS$_{\rm +III}$\\ 
\hline
$\sigma(\gamma)$ & 0.031 & 0.038\\
$\sigma(w_0)$ & 0.105 & 0.094\\
$\sigma(w_a)$ & 0.340 & 0.289\\
\end{tabular}
\caption{\label{pluspre}Gravitational growth index and dark energy 
equation of state uncertainties provided by each of the Stage IV 
experiments in conjunction with Stage III. } 
\end{table}
 
The complementarity of the other methods (supernova distances, CMB 
power spectra, and weak lensing shear) from Stage III in breaking 
degeneracies tightens the constraints on $\gamma$ produced by BigBOSS 
and JDEM-PS by a factor of 1.4.  Stage IV experiments using these 
techniques will further reduce the uncertainties on $\gamma$, either 
directly or indirectly through constraining other, correlated 
cosmological parameters. 

\begin{figure}
\begin{center}
\psfig{file=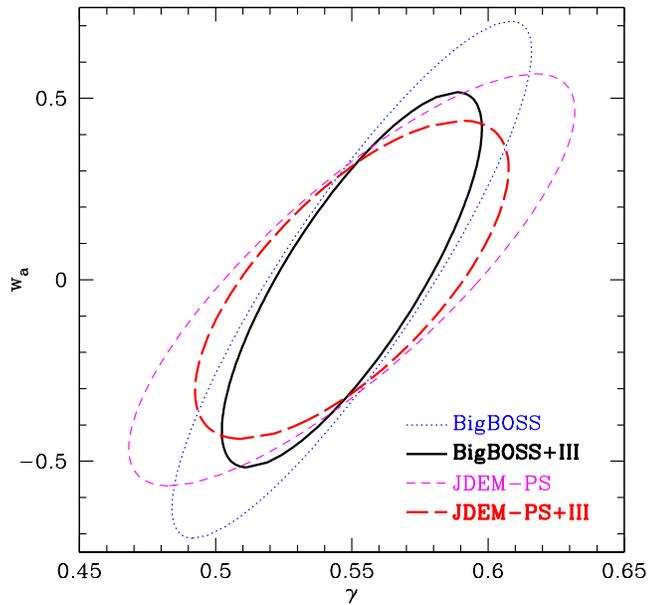,width=3.4in}
\caption{1$\sigma$ joint confidence contours for the gravitational 
growth index $\gamma$ and equation of state time variation $w_a$, 
marginalizing over the other parameters, are plotted for BigBOSS and 
JDEM-PS with and without Stage III information. 
}
\label{fig:gwa}
\end{center}
\end{figure}

BigBOSS from the ground and JDEM-PS from space appear comparable 
in their cosmology reach.  For the marginalized uncertainties, BigBOSS 
does better on the gravitational growth index $\gamma$ by a factor 1.26 
while JDEM-PS does better on the equation of state time variation $w_a$ 
by a factor 1.27.  We exhibit the joint 68\% confidence contours in 
Fig.~\ref{fig:gwa} where we see that the JDEM-PS contours are slightly 
fatter, having an overall area 1.23 times the BigBOSS constraints. 
Treating the inverse area of the parameter estimation contours as a 
figure of merit (FoM), Table~\ref{tab:fom} lists the ratios of FoM's 
for the BigBOSS plus Stage III and JDEM-PS plus Stage III experiments.

\begin{table}
\centering
\begin{tabular}{ccc}
 & BigBOSS/JDEM-PS & BigBOSS$_{\rm +III}$/JDEM-PS$_{\rm +III}$\\
\hline
$\gamma,\Omega_{\rm DE}$ & 0.93 & 0.99\\ 
$\gamma,w_0$ & 1.16 & 1.20\\ 
$\gamma,w_a$ & 1.21 & 1.23\\ 
$w_0,w_a$ & 0.88 & 0.86 
\end{tabular}
\caption{The ratios of the figures of merit (inverse areas) are given for 
various parameter spaces listed in the first column.  The second column 
shows the ratios for the Stage IV experiments alone; the third column 
includes Stage III information for each of them. 
}
\label{tab:fom} 
\end{table}

\section{Survey Characteristics}\label{ssurvey} 

In this section we investigate the influence of different survey 
parameters in the determination of the gravitational growth index $\gamma$. 
We discuss the influence of the redshift resolution $\sigma_z = R^{-1}$, 
the model for non-linear redshift distortions (i.e.\ the small scale 
velocities appearing in the Finger-of-God effect), the uncertainty in 
the bias parameter $b_{EL}$ shown in Table~\ref{spec}, and in 
particular the survey redshift range and design.  We will use the 
information coming from the full power 
spectrum as defined in Eq.~(\ref{eq:full}).  To clarify the effects we 
do not 
include information from Stage III experiments.

\subsection{Redshift Resolution} 

The effect of the uncertainty in the redshift measurement is incorporated by 
including a Gaussian suppression factor in the power spectrum in $k$-space: 
\beq
P_{\rm damp}(k,\mu) = P(k,\mu)\, e^{- k^2 \mu^2 \sigma_z^2 c^2/ H(z)^2}\,. 
\eeq
We include this factor in the Poissonian noise factor entering the 
effective volume, Eq.~(\ref{eq:veff}), but do not vary it with cosmology. 

A simple rule of thumb can be derived for the minimal resolution to 
achieve in order to neglect the influence of redshift uncertainties.  
Given that we truncate the integral defined in the Fisher matrix at 
$k_+ = 0.1\,h$/Mpc to exclude non-linear redshift distortions, this 
resolution effect will start to be significant when $k_\star\equiv 
H(z)/(c \sigma_z) \approx k_+$. This yields $\sigma_z \approx 0.003$ 
or $R \approx 300$. This is near the JDEM-PS minimal resolution, 
but this estimate is for the worst-case scenario ($k = k_+$, $\mu = 1$, 
$H(z)=H_0$), 
so redshift uncertainties should not be an issue for JDEM-PS/BigBOSS.  
For experiments with larger redshift measurement uncertainties, however, 
the effect on cosmology determination can be significant as shown for 
the full numerical computations in Table~\ref{res}.

\begin{table}
\centering
\begin{tabular}{cccc}
$R = \sigma_z^{-1}$ & $k_{\star,0}$ ($h$/Mpc) & BigBOSS & JDEM-PS\\
\hline
20 & 0.0067 & 0.110 & 0.124\\
200 & 0.067 & 0.044 & 0.054\\
$\infty$ & $\infty$ & 0.043 & 0.054
\end{tabular}
\caption{\label{res}Impact of the resolution $R$ on $\sigma(\gamma)$.}
\end{table}

As expected, the BigBOSS/JDEM-PS values $k_\star = 0.067 - 0.67\,h$/Mpc 
are sufficient for the cosmology estimation in our studies.  However, 
two issues must also be kept in mind: including information from 
$k>0.1\,h$/Mpc would increase the resolution requirements, and 
high resolution plays a key role in cleanly selecting the galaxy 
populations, e.g.\ avoiding line confusion in emission line galaxies.

\subsection{Non-linearities \label{snonlin}} 

We can examine how a better understanding of the transition to the 
non-linear part of the power spectrum could lead to an improvement in 
determining $\sigma(\gamma)$. Instead of truncating the integral in 
Eq.~(\ref{Fish}) entering the Fisher matrix at $k_+$, we can choose to 
implement a streaming model representing Fingers of God effects 
(see for example \citet{peeb}), where we 
integrate over all $k$ but multiply the power spectrum by a damping 
factor, either Lorentzian or Gaussian. This is supposed to model an 
exponential or Gaussian probability distribution function for the 
peculiar velocities of galaxies. We investigate the three forms of 
the small-scale velocity damping factors: 
\beqa
\textrm{Cutoff: }P_{nl}(k,\mu) & = & P(k,\mu)\, \Theta(k_+ - k)\\
\textrm{Gaussian: }P_{nl}(k,\mu) & = & P(k,\mu)\, e^{-(k/k_+)^2 \mu^2}\\
\textrm{Lorentzian: }P_{nl}(k,\mu) & = & \frac{P(k,\mu)}{1 + (k/k_+)^2 \mu^2}
\eeqa
where $\Theta$ is the Heaviside function. 
We use $k_+ = 0.1\,h$/Mpc for all cases.  
Table~\ref{nonlinea} shows the effects on the determination of the 
gravitational growth index.

\begin{table}
\centering
\begin{tabular}{ccc}
Case & BigBOSS & JDEM-PS\\
\hline
Cutoff & 0.043 & 0.054\\
Gaussian & 0.024 & 0.026\\
Lorentzian & 0.019 & 0.021\\
\end{tabular}
\caption{\label{nonlinea}The impact of different models for the translinear 
damping due to peculiar velocities on the gravitational growth index 
estimation $\sigma(\gamma)$.  To restrict to the translinear scales we 
further truncate the power spectrum integral at $k = 1\,h$/Mpc.} 
\end{table}

We see that the statistical uncertainty on $\gamma$ is largest if we 
simply cut out all translinear information, by about a factor 2.  Thus 
we have adopted the most conservative method to predict $\sigma(\gamma)$; 
the information might not be completely lost on translinear scales, but 
only attenuated by Finger-of-God effects.  Adopting a Gaussian or 
Lorentzian damping model allows extraction of some information, with the 
choice of model affecting the results at the $\sim25\%$ level. 
However, an exponential or Gaussian probability distribution function for 
the streaming model is still not completely accurate, and along with the 
reduced statistical uncertainty on $\gamma$ could come a systematic bias.  
Thus we retain the conservative, cutoff method. Taking into account a 
halo model, for example \citet{halo}, could allow a more detailed 
investigation of the proper treatment of the translinear regime. 
Issues of nonlinear bias could also arise beyond the 
$k_+=0.1\,h$/Mpc adopted in this paper.

\subsection{Redshift Range and Survey Design} 

While the survey volume due to the solid angle $\Omega_{\rm sky}$ simply 
scales the parameter estimation as $\sigma(\gamma) \propto 1 / 
\sqrt{\Omega_{\rm sky}}$ in the statistical treatment without priors, the 
influence of redshift range is more complex and interesting.   Since 
the galaxy population used also depends on redshift we simultaneously 
investigate the influence of the galaxy bias values. 

Table~\ref{pop} shows the results for considering the populations, 
and their associated redshift ranges, one at a time and also in combination 
with different values (0.8 vs.\ 1.2) for the emission line galaxy 
population bias.

\begin{table}
\centering
\begin{tabular}{ccc}
Populations & BigBOSS & JDEM-PS\\
\hline
LRG, $b_{LRG} = 1.7$ & 0.067 & 0.115\\
EL, $b_{EL} = 0.8$ & 0.574 & 0.187\\
EL, $b_{EL} = 1.2$ & 0.503 & 0.197\\
LRG + EL, $b_{EL} = 0.8$ & 0.043 & 0.054\\
LRG + EL, $b_{EL} = 1.2$ & 0.042 & 0.053\\
\end{tabular}
\caption{\label{pop}Impact of the redshift range and the associated two 
different populations on $\sigma(\gamma)$.  The top three lines consider 
a single population and its redshift range from Table~\ref{spec}, while 
the bottom two lines combine both populations and their redshift ranges. 
The second vs.\ third, and fourth vs.\ fifth, lines examine the effect 
of different values for $b_{EL}$.}
\end{table}

As found in \citet{linder-gamma}, most of the constraint on $\gamma$ comes 
from the redshift range $z\lesssim1$, which mostly corresponds to the LRG 
population.  The reason is simple: the cosmological information on 
$\gamma$ enters the power spectrum through the factor $\Omega_m(z)^\gamma$, 
so at higher redshifts where $\Omega_m(z)$ is closer to $1$, the 
sensitivity to $\gamma$ decreases.  
The value of the EL bias adopted does not have a significant effect, 
especially when in combination with the low redshift, LRG sample.  
Furthermore, note that the EL only case for JDEM-PS, which includes all 
the information from JDEM-PS itself 
and none of the data to be provided by BOSS, only determines 
$\sigma(\gamma)\approx0.2$, even though the sample extends down to $z=0.7$. 
For JDEM-PS, the BOSS data enable an improvement of almost a factor 4 in 
the growth index parameter determination. 

These consequences of redshift range raise an important question: what is 
the science reach of the BigBOSS survey if the EL sample 
is shifted from $z=1-2$ to $z=0.7-1.7$?  This not only changes the 
redshift range of the EL sample information but creates an overlap 
between LRG and EL information.  The generalization of Eq.~(\ref{Fish}) 
to multiple galaxy populations \citep{urossamp,white} reads: 
\beq
F_{ij} = \sum_{XY} \frac{V_0}{2(2 \pi)^3} \int d^3 k  \frac{\partial P_X}{\partial p_i} C_{XY}^{-1} \frac{\partial P_Y}{\partial p_j} ,
\eeq
where X and Y are indices describing pairs of galaxy populations, and 
$C_{XY}$ is the 
covariance matrix of the power spectra. Adapting the BigBOSS specifications 
from Table~\ref{spec} by shifting the EL sample to $z=0.7-1.7$ retains 
the science leverage and in fact delivers a mild improvement of 8\%: 
\beqa 
\textrm{BigBOSS standard}&:& \sigma(\gamma)=0.043 \\ 
\textrm{BigBOSS }z_{EL}=0.7-1.7&:& \sigma(\gamma)=0.040 
\eeqa 

Moreover, 
a redshift maximum of 1.7 reduces the technical complexity of the data 
acquisition and analysis, greatly ameliorating issues of line confusion 
and reduced signal-to-noise that occur over $z=1.7-2$.  (Note that for 
$z>2$ Ly-$\alpha$ enters the spectral range and the issues again 
disappear.)  The overlap of LRG and EL populations with very different 
biases in the same redshift range $z=0.7-1.0$ also offers the possibility 
of crosscorrelation and reduction of sample variance \citep{urossamp}.  Thus, 
these results motivate shifting the EL 
redshift range to $z=0.7-1.7$, achieving $\sigma(\gamma)=0.040$ (and 
0.030 with Stage III information).

\subsection{Galaxy Samples} 

The values of the galaxy number densities and biases listed in 
Table~\ref{spec} come from the references given.  While it is beyond 
the scope of this paper to do detailed survey design, we can explore 
whether some variations in the adopted values matter.  

As we have seen in 
the previous subsection, a change in the constant bias of the ELG 
population from 0.8 to 1.2 has a 2\% effect on determining 
$\gamma$.  We now consider an evolving model for bias.  Motivated by 
\citet{padma06} we take $b_{LRG}=b_1+0.4z$ with fiducial $b_1=1.6$, 
and motivated by \citet{sumiyoshi} we take $b_{EL}=b_2+(z-0.7)/2.6$ 
with fiducial $b_2=1$.  

The constraints on the dark energy parameters $\gamma$, $w_0$, $w_a$ 
improve by 4\%, 2\%, 6\% for BigBOSS and degrade by 5\%, 11\%, 10\% 
for JDEM-PS.  These changes are due to altered covariances between the 
bias parameters and the dark energy parameters, involving an interplay 
between the $nP$ factor in the effective volume and the Fisher 
sensitivity $\partial\ln P/\partial b_i$.  Note that the latter 
quantity goes as $2/(b_i+f\mu)$, so an increased bias decreases the 
Fisher element.  However, increasing the bias increases the effective 
volume through raising $nP$.  In the BigBOSS case, this second factor 
is more than sufficient to compensate for the reduced sensitivity. 
However, JDEM-PS has such a high galaxy number density that the change 
in $nP$ has little effect on the effective volume, leaving only the 
reduced sensitivity.  Updating Table~\ref{tab:fom} for the evolving 
bias case, Table~\ref{tab:fombz} shows further gains in the figures 
of merit for BigBOSS relative to JDEM-PS.

\begin{table}
\centering
\begin{tabular}{ccc}
 & BigBOSS/JDEM-PS & BigBOSS$_{\rm +III}$/JDEM-PS$_{\rm +III}$\\
\hline
$\gamma,\Omega_{\rm DE}$ & 1.19 & 1.12\\ 
$\gamma,w_0$ & 1.45 & 1.41\\ 
$\gamma,w_a$ & 1.52 & 1.42\\ 
$w_0,w_a$ & 1.13 & 1.07 
\end{tabular}
\caption{As Table~\ref{tab:fom} but using an evolving bias model 
for the galaxy populations.  The ratios of the figures of merit 
(inverse areas) are given for 
various parameter spaces listed in the first column.  The second column 
shows the ratios for the Stage IV experiments alone; the third column 
includes Stage III information for each of them. 
}
\label{tab:fombz} 
\end{table}

Regarding the number densities used for the galaxy populations, these 
come from selection functions of the survey with respect to the intrinsic 
populations within the detection limits.  In general, target selection 
is a complicated procedure and these numbers represent a sculpted 
target sample not a flux- or volume-limited distribution.  We consider 
one simple variation in the BigBOSS ELG distribution, motivated by 
the previous subsection where the redshift range was shifted from 
$z=1-2$ to $z=0.7-1.7$.  Such a shift was found to slightly improve 
the cosmology constraints, and it also reduces the amount of time needed 
to observe the galaxies.  If we take advantage of this by now looking at 
a survey plan with four times the number density of ELG in the range 
$z=0.7-1$ (and $z=1-1.7$ unchanged), we find further improvements in 
determination 
of $\gamma$, $w_0$, $w_a$ by 2\%, 4\%, 3\% relative to the uniform 
number density in $z=0.7-1.7$ case of the previous subsection. 

These calculations show that the basic point of ground and space 
surveys being capable of delivering comparable cosmology constraints is not 
very sensitive to these variations in the survey design.  Detailed 
experiment design and optimization, however, is beyond the scope of 
this paper. 
We have not considered other experiments such as the Euclid space 
mission \citep{euclid}, since it includes other cosmological probes on 
a par with the power spectrum measurement, and 21 cm mapping 
surveys such as SKA (see \citet{peterson,morales}), since neutral 
hydrogen gas measurement techniques and precision constraints are 
not as fully developed.

\section{Conclusions}\label{sconcl}

The three-dimensional distribution of large scale structure contains 
information on both the cosmological parameters and testing gravity. 
We have studied the capabilities of next-generation power spectrum 
experiments from the ground, BigBOSS, and from space, JDEM-PS, to use 
the baryon acoustic oscillations, power spectrum shape, and redshift 
space distortions to test standard cosmology. 

The main conclusion is that the two experiments could achieve comparable 
constraints.  We emphasized the importance of including simultaneously 
the parameters 
that affect growth -- the gravitational growth index characterizing 
deviations from general relativity, the dark energy equation of state 
value and its time variation, and neutrino mass.  Including these and 
other cosmological parameters we estimate the uncertainty on determination 
of the gravitational growth index to be 0.043 for BigBOSS, 0.054 for JDEM-PS, 
or 0.031 and 0.038 respectively when combined with nearer-term, Stage III 
experiments.  This represents nearly an order of magnitude improvement 
over Stage III knowledge. 

We have also studied the survey characteristics and confirm that 
the power spectrum at redshifts $z\lesssim1$ has strong leverage.  
This makes the luminous 
red galaxy component of the survey quite important.  Furthermore, our 
results demonstrate that shifting the redshift range of the emission 
line galaxy survey of BigBOSS from $z=1-2$ to $z=0.7-1.7$ can improve 
the constraints, while adding benefits such as reduced technical 
complexity and line confusion and increased signal-to-noise and the 
ability to crosscorrelate galaxy populations of different biases. 

Lyman-$\alpha$ forest spectra from BigBOSS quasars at $z>2$, which 
we have neglected, will further advance the determination of 
cosmological parameters. 

The prospects for testing standard cosmology and in particular general  
relativity are promising.  Improved understanding of the translinear 
density regime and velocities would further extend the number of usable 
power spectrum modes, while complementarity with other Stage IV experiments 
utilizing supernova distances, CMB measurements, and weak lensing data 
would give powerful leverage on both the gravitational growth index and 
other cosmological parameters.  The capability of probing beyond-Einstein 
gravity opens up a new window for our understanding of cosmic acceleration 
and fundamental physics.

\section*{Acknowledgments}

AS thanks N. Mostek, N. Padmanabhan and D. Schlegel for various 
insights about BigBOSS and JDEM and R. de Putter for 
valuable help on 
CMB\-easy.  EL thanks M. White and gratefully acknowledges support 
from a Chaire Blaise Pascal 
grant and hospitality from LPNHE and APC Paris.  
This work has been supported in part by the
Director, Office of Science, Office of High Energy Physics, of the
U.S.\ Department of Energy under Contract No.\ DE-AC02-05CH11231.

\label{lastpage}
\end{document}